# Lepton and Meson Masses

**Gustavo González-Martín**

Departamento de Física, Universidad Simón Bolívar,

Apartado 89000, Caracas 1080-A, Venezuela.

Webpage: http://prof.usb.ve/ggonzalm/

The lepton mass ratios are calculated using a geometric unified theory, taking the leptons as the only three possible families of **topological excitations** of the electron or the neutrino. The theoretical results give 107.5916 Mev for the **muon** mass using $m_\mu/m_e$ and 1770.3 Mev for the **tau** mass using $m_\tau/m_\mu$. Using the additional geometric interaction energy in a muon-neutrino system, the main leptonic mass contribution to the **pion** and **kaon** mass is calculated to be, respectively, 140.88 Mev and 494.76 Mev. The necessary first order corrections, due to the interaction of the excitations, should be of the order of the discrepancies with experimental values. The three geometric families of leptonic excitations may be related to a **quark** structure.

PACS :04.50.+h; 14.60.-z; 02.20.+h

# 1. Introduction.

We have presented a definition of mass [1 ], within a geometric relativistic unified theory of gravitation and other interactions [2 ], in terms of the concept of self energy of the non linear self interaction in a geometric space. This determines the particle mass in Dirac's equation as a representation of the structure group. The microscopic physical objects (geometric particles) are realized as linear geometric excitations, geometrically described in a jet bundle formalism shown to lead to the standard quantum field theory techniques. These geometric excitations are essentially perturbations around a non-linear geometric background space solution, where the excitations may be considered to evolve with time. The background space carries the universal inertial properties which should be consistent with the ideas of Mach [3] and Einstein [4 ] that assign fundamental importance of far-away matter in determining the inertial properties of local matter including the inertial mass.

The geometry is related to a connection $\Gamma$ in a principal fiber bundle $(E,M,G)$ and matter is related to a current $J$ of geometrical objects. The structure group $G$ is SL(4,R) and the even subgroup $G_+$ is $SL_1(2,C)$. The subgroup $L$ (Lorentz) is the subgroup of $G_+$ with real determinant, in other words, SL(2,C). There is only another subgroup $P$ in the possible group chains $G\supset H\supset L$, which is Sp(4,R). The two coset $G/G_+$ y $P/L$ determine symmetric spaces that we shall denote respectively by $K$ and $C$. These groups have a principal fiber bundle structure over the cosets, indicated as $(G,K,G_+)$ and $(P,C,L)$. The induced representations are carried by a space $D$, which is an associated bundle to $G$ with Lorentz group representations $\mathcal{D}[L]$ as fiber. The energy density of induced representations over the symmetric spaces is expressed by the product $J\bullet\Gamma$. The masses are obtained integrating over the symmetric coset subspaces of relativistically inequivalent points $K_R\subset K$ for the group $G$ and $C_R\subset C\subset K$ for the subgroup $P$,

$$\int J \bullet \Gamma dk \quad . \qquad (1)$$

Due to the existence a constant non-linear background solution, called the substratum, the quotient of the masses for these geometric $G$-excitations and $P$-excitations around the substratum may be expressed in terms of volumes associated to these spaces and have the exact finite value (see appendix 6.4),

$$\frac{m_G}{m_P} = \frac{V(K_R)}{V(C_R)} = 6\pi^5 = 1836.1181 \approx \frac{m_p}{m_e} \quad , \qquad (2)$$

which is a good approximation for the quotient of the experimental physical values for the proton and electron masses. The geometric expression for this mass quotient was known [5, 6 ,7 ], but not physically explained.

Now we present the application of these ideas to the calculation of the masses of topological excitations that may represent heavy leptons. Preliminay partial results were presented at the III LASSF (III Congreso Ven. de Física) [8].

# 2. Families of Scattering Solutions.

Now consider only topological properties, independent of the connection, of the space of complete solutions (substratum plus excitation solutions). An incoming scattering solution is a jet bundle local section, over a world tube in the space time base manifold, that describes the evolution of the solution in terms a time like parameter $\tau$ from past infinity to some finite time $t$. Similarly, an outgoing solution is a local section from time $t$ to future infinity. The local sections in the bundle represent classes of solutions relative to local observers. Scattering solutions at infinity are asymptotically free excitation solutions around a substratum. The substrata (incoming and outgoing) are equivalent to each other and to the constant substratum solution if we choose observer frames adapted to the substrata.

Since the equations are of hyperbolic type, we should provide initial conditions on an initial tridimensional hypersurface at past infinity $I_-$. We require that all incoming solutions, at the past infinity hypersurface $I_-$, reduce to a free excitation around the substratum solution at the bidimensional spatial infinity subspace $I_-(\infty)$. Since the incoming solution substrata are equivalent to the substratum solution at spatial infinity $I_-(\infty)$, we may treat this spatial infinity as a single point, thus realizing a single point compactification of $I_-$, so the initial hypersurface $I_-$ is homeomorphic to $S^3$. All incoming solutions on $I_-$ are classified by the functions over $S^3$. The same requirements may be applied to the outgoing remote future solutions and, in fact, to any solution $I$ along an intermediate tridimensional hypersurface, a section of the world tube. Thus, the final hypersurface at future infinity $I_+$ is also homeomorphic to $S^3$. The incoming and outgoing substratum local sections over $I_-$ and $I_+$ must be pasted together in some common region $I\times R$ around the present $t$, by the transition functions of the bundle. The scattering interaction is represented by the group action of the transition functions at $\tau=0$. All generators of the group produce a transformation to a different, but equivalent under the group, expression for the solution. If the holonomy group of the solution is not the whole group, there is a reference frame that reduces the structure group to the particular holonomy subgroup. But in general for arbitrary observers, there are solutions formally generated by SL(2,*Q*)=SL(4,R). Since this transition region, the “compactified equator”×*R*, has the topology of $S^3 \times R$, the transition functions $\varphi$ define a mapping, at the $\tau=0$ hypersurface,

$$\varphi : S^3 \to SL(4,R) \qquad (3)$$

which is classified by the third homotopy group [9 ] of the structure group SL(4,R) or the respective holonomy subgroup. There are some solutions not deformable to the trivial solution by a homeomorphism because $\varphi$ represents the twisting of local pieces of the bundle when glued together.

**These scattering solutions are characterized by integer topological numbers related to the maximal compact subgroup, known as winding or wrapping numbers.** In all cases the scattering solutions are characterized by a topological number *n* and in particular the scattering solutions of *G*-excitations have an additional topological number *n'*. This result implies that there are solutions $\Gamma_n$, $J_n$ that are not homotopically equivalent to $\Gamma_0$, $J_0$.

## 3. Geometric Excitation Masses

The integrand $J \bullet \Gamma$, which corresponds to a fundamental representation, is a local substratum constant equal for all values of *n*. The number of possible states of a fundamental representation only depends on the integration volume. For *P*-excitations, when integrating over the $C_R$ subspace of De Sitter space *C*, corresponding to inequivalent observer states by an *L* transformation, we have counted the number of *L* equivalence classes of states (points in the mass hyperboloid are in the same relativistic equivalence class as the rest state of the representation) in local form (only counting $n=0$ local states). Taking in consideration the global characteristics of the topological excitation that would represent heavy leptons, the count should be larger, over all *L* equivalence classes of excitation states with wrapping *n*. What is the number states for each value of *n*?

Since we are working with a manifold with atlas, the integration equation (1) over the symmetric space *K* is realized en the atlas charts. On the principal fiber bundle (*E*,*M*,*G*), the transition function preserves the projection and acts, as a group element, over the fiber which is the bundle structure group *G* (see appendix 6.1). For two neighborhoods *U*, *V* in the base manifold *M*, which corresponds to space-time, we have the following maps: homeomorphisms *h* from the bundle to the model space, transition functions $\varphi$ among the charts and partial identity *i* over the points $m \in M$,

$$\begin{array}{ccc} U(E) & \xrightarrow{h_U} & U(M)\times G \\ \downarrow i_V^U & & \downarrow \varphi_V^U \\ V(E) & \xrightarrow{h_V} & V(M)\times G \end{array} \quad . \qquad (4)$$

The transition function $\varphi$ acts as an element of *G* over the incoming scattering solution, defined on a hypersurface homeomorphic to $S^3$, to produce the outgoing scattering solution. Physically $\varphi$ represents a collision interaction. The active representation is the interaction of two excitations observed by a single observer. The passive representation is the observation of a single excitation by different observers. The non trivial transition functions of class $[\varphi_n]$, where the index *n* represents the set of wrapping numbers (*n*, *n'*), are classified by the mappings (see appendix 6.2)

$$S^3 \to SL(4,R) = G \quad . \qquad (5)$$

Consider *G* as a principal fiber bundle $(G,K,G_+)$. The induced representations, assigned to points $m \in M$, are sections of an associated bundle to *G* that has Lorentz group representations $\mathcal{D}[L]$ as fiber and is denoted by $(D,K,\mathcal{D}[L],L)$. Under a change of charts, representing a change of observers, the transition function acts on the bundle *D*, on the integrand in equation (1), a section $J \bullet \Gamma$ over *K*, which we shall indicate by *f*(*m*).

The transition functions belong to the different classes determined by the third homotopy group of *G* (see appendix 6.3). The class $[\varphi_n]$ may be expressed, using the homotopic product in terms of the class $[\varphi_1]$

$$[\varphi_n] = [\varphi_0] \circ [\varphi_1]_1 \circ [\varphi_1]_2 \circ \cdots [\varphi_1]_n = [\varphi_0] \circ [\varphi_1]^n \quad . \qquad (6)$$

This class $[\varphi_n]$ may be considered generated by the product of *n* independent elements of class $[\varphi_1]$ and one element of class $[\varphi_0]$. Observe that the homotopic mapping

$$h : S^3 \to (G,K,L) \qquad (7)$$

determines essentially a wrapping of subspaces $s(m) \subset G$, wrappings homeomorphic to spheres $S^3$, inside $(G,K,G_+)$. Each of the generating classes $\varphi$ is globally associated to a generating wrapped subspace $s$ in $(G,K,G_+)$, additional and necessary, linked to the original trivial subspace by a transition function. In this manner the class $[\varphi_n]$ is associated to the trivial wrapping $[s_0]$ and to $(n, n')$ additional wrappings $[s_1]$, defined by the expression for the coordinates $u$ of point $m \in M$

$$\varphi_{iV}^{U}\left(u(m), s(u(m))\right) = \left(v(m), s_i(v(m))\right) \quad . \tag{8}$$

The excitation is not completely described by a single subspace that would correspond to a single observer $s(m)$ in the passive representation. For global reasons we must accept the presence of as many images of generating subspaces in the non trivial chart as there are wrappings. Nevertheless some of the wrappings may be equivalent under the subgroup of interest. For the trivial subbundle $G$ there is only one independent wrapping, the trivial $(n=0)$ wrapping $s_0$, because the subspaces $s_n$ are equivalent under a $G$ transformation. For other subbundles, a tridimensional su(2) subalgebra acts on the complete group $G$ fiber bundle, generating transformations of $P$ and $L$ subbundles which are not equivalent under the corresponding subgroup. It is possible to generate no more than three wrappings $s_i(m)$ in geometrically independent $P_i$, $L_i$ subbundles in $G$. In other words any other wrapping may be obtained by some combination of three transformations in $G$. For this reason the number of independent wrappings $s_i$ in $P$ or $L$ bundles is $n \leq 2$ and determines three families of excitations.

## 4. Leptonic Masses.

The actions of the $n+1$ subspaces $s_i(m) \subset P_i$ $(0 \leq i \leq n)$ on a Lorentz representation over $C$, the section $f \in \mathcal{D}$ of fiber bundle $(D,C,\mathcal{D}[L],L)$,

$$f(m): C \to D \tag{9}$$

map the fiber $\pi^{-1}$ over $c$ onto fibers $\pi^{-1}$ over $s_i c$ by linear transformations $l_i$ [10 ],

$$s_i(m): \pi^{-1}(c(m)) \to \pi^{-1}(s_i(m)c(m)) \quad , \tag{10}$$

$$s_i(m)\left(f(m, c(m))\right) = l_i(s_i(m))\, f(m, s_i(m)c(m)) \quad , \tag{11}$$

that determine a set of $n+1$ images of $f$, independent wrapped sections $f_i$ in the non trivial chart, one for each independent wrapping $s_i$,

$$f_i(m): C \to D \quad , \tag{12}$$

$$f_i(s_i c) \equiv l_i(s_i)\, f(s_i c) \quad . \tag{13}$$

The Lorentz representations $f_i(s_i c)$ may be independent of any other $f_a(s_a c)$. Therefore, the physically possible states of a wrapped excitation of class $[\varphi_n]$ correspond 1 to 1 to all $L$ equivalence classes of states determined in each one of these $n+1$ section images $f_i$. Using the passive representation, on each image there are momentum coordinates $k_i^{\mu}$ relative to corresponding local observers $i$, states physically independent among themselves. The integration should be done over all independent variables $k_0^{\mu}, k_1^{\mu}, \ldots k_n^{\mu}$ , that is over the set of classes $C_R$ of relativistic equivalent spaces contained in a product of $n+1$ De Sitter spaces $C$.

As we indicated (see appendix 6.4) the integrand $J \bullet \Gamma$, corresponding to a fundamental representation, is a local constant equal for all values of $n$. The bare masses are determined by the volume of integration. Let us denote the integration space that supports $f_i(m)$ by $C^*$, equal to product of $n+1$ copies of the original $C_R$ space, corresponding to the $n+1$ observed wrappings

$$C^* = (C_R)^{n+1} \quad . \tag{14}$$

For the trivial case $(n=0)$ the calculation is strictly local, it is not necessary to consider wrapped charts and the integration over $C_R$ may be done locally on any chart of the trivial class. The subspace $s_0$ is determined by a local observer.

For wrapped states $(n \neq 0)$ we are forced to use the non trivial transition functions $[\varphi_n]$, with non zero wrapping $n$ $(n \neq 0)$, that relates a wrapped chart of class $[n]$ with the trivial chart. Therefore it is necessary to consider simultaneously (globally) these pairs of chart classes associated to the incoming and outgoing scattering solutions respectively. Physically it is necessary to consider all $L$-equivalence classes of possible states in accordance with observers related to non trivial charts obtained from the trivial chart. Therefore, for the class $[n \neq 0]$ it is necessary to integrate over all possible charts with "globally wrapped" sections, produced by the action of transition functions. In this case the transition functions may be any element of the group *P*. We restrict the integration to those non trivial charts that are inequivalent under a relativity transformation *L* and the number of possible charts corresponds to the volume of the *P/L* subspace $C_R$ of *L*-equivalence classes.

The linear topological excitations are generated by the linear infinitesimal action of the differential $\varphi_*$ of the transition function $\varphi$ that represents the $\tau=0$ interaction. When restricting the transition functions to elements of the subgroup *P*,

$$V \xrightarrow{h_U} A(V) = V \times P \quad , \qquad (15)$$

we should only consider the effect of the differential mapping $\varphi_*$. The odd subspace in the *P* algebra, as differential mapping, generates the coset *C*. The only generator inequivalent to an *L* "boost" is the compact u(1) generator. The differential action produces a smaller space of *L*-equivalence classes. If we denote the compact subgroup U(1) ⊗ SU(2) of *P* by *H* we obtain the non compact coset $B \supset C$

$$B = \frac{P}{H} \quad . \qquad (16)$$

The group *P* may be expressed as a principal fiber bundle (*P*,*B*,*H*) over *B*. If we inject *L* in *P* the image of the rotation subgroup in *L* should be the SU(2) subgroup in *H* but the image of the "boost" sector in *L* is not uniquely defined in *B*. The group *L* acts on *P* preserving its image in *P*. The u(1) subalgebra of *H* acts on *B*, as a translation within *P*, mapping the *C* subspace onto another subspace *C'* in *B* which is *P*-equivalent to *C*. Nevertheless *C´* is not *L*-equivalent to *C*. Therefore, the possible *L*-inequivalent states correspond to the translation action by the U(1) compact subgroup not related with the rotation SU(2). The number of *L*-equivalence classes of these possible charts, over which we should integrate, is determined by the volume of this electromagnetic U(1) group. The total wrapped space $C^T$ for an *n*-excitation ($n\neq 0$) is the space formed by all possible translated $C^*$ spaces, not relativistically equivalent

$$C^T = U(1) \times C^* \quad . \qquad (17)$$

The number of states (different *k* values) is proportional to the volume of this total $C^T$ space, which may be calculated,

$$V(C^T) = V(U(1)) \times \left(V(C_R)^{n+1}\right) \qquad n \neq 0 \quad . \qquad (18)$$

The bare mass of the trivial excitation ($n=0$) which is the electron, as previously indicated, is proportional to the volume of $C_R$ (see appendix 6.4)

$$V(C_R) = \frac{16\pi}{3} \quad . \qquad (19)$$

The bare masses corresponding to the *n*-excitations ($n\neq 0$) are proportional to the volumes

$$V(C^T) = 4\pi \left(\frac{16\pi}{3}\right)^{n+1} \qquad 0 < n \le 2 \quad , \qquad (20)$$

that may be expressed in terms of the electron bare mass $m_0$,

$$m_n = m_0 \left(\frac{16\pi}{3}\right)^n 4\pi \qquad 0 < n \le 2 \quad . \qquad (21)$$

For excitations with wrappings 1 and 2 we have

$$\frac{m_2}{m_1} = V(C_R) = \frac{16\pi}{3} = 16.75516 \approx \frac{m_\tau}{m_\mu} = 16.818 \quad , \qquad (22)$$

$$\begin{aligned} m_1 = m_0 \left(\frac{16\pi}{3}\right) 4\pi &= 0.5109989 \times 210.5516 \\ &= 107.5916 \ \text{Mev} \approx m_\mu + O(\alpha) \quad , \end{aligned} \qquad (23)$$

$$m_2 = m_1\left(\frac{16\pi}{3}\right) \approx m_\mu\left(\frac{16\pi}{3}\right) = 105.6584 \times 16.75516$$
$$= 1770.3 \text{ Mev} \approx m_\tau + O(\alpha) \quad , \tag{24}$$

corresponding to the **bare masses of the $\mu$ and $\tau$ leptons**. Due to the electron-muon U(1) electromagnetic interaction in the topological excitation, we should apply a first order energy correction to $m_1$. For the subbundle $L$ we can make a similar calculation. The bare masses of the associated neutrinos for the 3 families are zero because the volume of the $L/L$ coset space is zero. We shall refer to these topological excitations, respectively, as $T_nP$-excitations and $T_nL$-excitations.

### 4.1. Masked Excitations.

The chart transformations under general $G$ transition functions translate, within $G$, the $C_R$ region of integration. The only possible additional $L$-inequivalent states correspond to the action of compact sectors in the group. In order to find the different possibilities we consider the two related chains of symmetric spaces in $G$, not related to $L$

$$\begin{array}{ccccccc} G & \supset & K_R & \supset & C_R & \supset & I \\ \cup & & \cup & & \cup & & \\ SU(2) & \supset & SU(2)/U(1) & \supset & U(1) & \supset & I \end{array} \quad . \tag{25}$$

The compact symmetric subspaces $C_R$ and $K_R$, respectively, contain one and two U(1) electromagnetic subgroups of the three equivalent U(1) subgroups contained in the electromagnetic SU(2). We note that one U(1) is common to both subspaces. We are interested in extending the translation within $G$, beyond the U(1) transition function, by adjoining compact sectors. The physical interpretation is that these excitations are subjected to additional nonlinear interactions that increase the energy and therefore their masses. In the same manner as an observable relativistic motion increases the rest mass to a kinetic mass, an observable relativistic interaction increases the free mass to a dynamic mass. This relativistic effect, included in the mass definition, is realized by the action of the transition functions at $\tau=0$.

For $n\neq0$ topological excitations the transition functions may be extended from U(1) to $C_R$. The physical interpretation is that these topological excitations are subjected to the additional electroweak interactions of a leptonic $P$-system. We say that the excitations are "masked" by the nonlinear electroweak interactions. The geometric mass expressions for the class [$n\neq0$] masked excitations should be multiplied by the volume of the corresponding complete subspace of $L$-equivalent classes $V(C_R)$ rather than by its $V$(U(1)) subspace. The weak interaction energy acquired by the $n$-excitations corresponds to the product of the values from equations (23) and (24) by the ratio of the volumes

$$\frac{m_1'}{m_1} = \frac{V(C_R)}{V(U(1))} = \frac{4}{3} \approx \frac{m_\pi}{m_\mu} = 1.320957 \quad . \tag{26}$$

If we use experimental values for the lepton masses, we obtain the masked geometric masses,

$$m_1' = 140.8778 \text{ Mev} \approx m_\pi + O(\alpha) \; , \tag{27}$$

$$m_2' = 2369 \text{ Mev} \approx m_f + O(\alpha) \; . \tag{28}$$

We shall refer to these $C_R$ masked (electroweakly interacting) topological excitations as $T_nP_C$-excitations. The $T_nP_C$-excitations may be considered components of mesons. In particular, a masked muon $\mu$' or $T_1P_C$-excitation, joined to a low energy $T_1L$-excitation, has the geometric mass and other properties of the **pion** $\pi$.

For $n\neq0$ topological excitations the transition functions may be further extended beyond $C_R$ to include the sector of $K_R$ corresponding to both U(1) subgroups in $K_R$ which is an SU(2)/U(1) compact sector that may be identified with the sphere $S^2$. The physical interpretation is that these excitations are subjected to additional strong nonlinear interactions of a hadronic $G$-system. The parametrization of group spaces and their symmetric cosets is, to a certain extent, arbitrary. They map different points in the linear Lie algebra to the same group operation. Since both equivalent U(1) subgroups in $S^2$ have the same significance because of the symmetry of $S^2$, they both should contribute equally to the invariant volume of $S^2$ and it is convenient to choose a parametrization that explicitly displays this fact. A parametrization that accomplishes this is

$$SU(2) = \exp(\alpha_+ J_+) \exp(\alpha_3 J_3) \exp(\alpha_- J_-) = \Sigma_+ U(1) \Sigma_- \quad , \tag{29}$$

$$V(SU(2)) = 16\pi^2 = V(\Sigma_+) V(U(1)) V(\Sigma_-) = 4\pi V^2(\Sigma_+) \quad . \tag{30}$$

We adjoin the group subspace $\Sigma \supset K_R$, defined by the expression in equation (29), to subspace $C_R$. Because of the extended integration in $\Sigma$, the $n$-excitations acquire a strong interaction energy corresponding to the ratio of the volumes

$$\frac{m_1''}{m_1'} = \frac{V(\Sigma C_R)}{V(C_R)} = V(\Sigma) = (4\pi)^{\frac{1}{2}} = 3.5449 \approx \frac{m_K}{m_\pi} = 3.5371 \quad . \tag{31}$$

We obtain the geometric masses for $n=1$ and $n=2$

$$m_1'' = m_\pi V(\Sigma) = m_\pi (4\pi)^{\frac{1}{2}} = 494.76 \quad \text{Mev} \; = m_K + O(\alpha) \quad , \tag{32}$$

$$m_2'' = m_2' V(\Sigma) = \left(\frac{16\pi}{3}\right) m_1' (4\pi)^{\frac{1}{2}} \approx \left(\frac{m_\tau}{m_\mu}\right) m_K = 8303 \quad \text{Mev} \quad . \tag{33}$$

These excitations are not contained in the $P$ sector defined by the $n=0$ wrapping, but rather in a combination of inequivalent $P$ sectors inside $G$. Therefore, strictly, they do not have a proper constituent topological $P$-excitation. We shall refer to these $\Sigma$ masked (strongly interacting) topological excitations as $T_n P_\Sigma$-excitations. In particular, a masked muon $\mu$” or $T_1 P_\Sigma$-excitation, joined to a low energy $T_1 L$-excitation, has the geometric mass and other properties of the **kaon** $K$.

The physical interpretation of these masked leptons suggests that the corresponding geometric excitations may be combined to form lepton pairs that may be considered particles. In accordance with this interpretation, the combinations under nuclear interactions are only possible if there is, at least, one masked lepton. Take the $n=1$ topological leptons and construct a doublet $l$ of a $^c$SU(2) combinatory group associating a masked muon and a stable lepton. We should indicate that the conjugation in the Clifford algebra, which is equivalent to the dual operation in sp(4,R), is not the dual operation in $^c$su(2). Both $l$ and its conjugate are $^c$su(2) fundamental representations **2**. The product is

$$\mathbf{2} \otimes \mathbf{2} = \mathbf{3} \oplus \mathbf{1} \quad . \tag{34}$$

The first possibility is that $\mu$ is a weakly masked $\mu$’, part of the SU(2) leptonic system doublet ($\mu$’, $\nu$) characterized by the charge $Q$ and the muonic number $L_\mu$. We get the $\pi$ representation:

$$\begin{aligned} [\mu', \nu] \otimes [\bar{\mu}', \bar{\nu}] &\equiv [\pi^-, \pi^0, \pi^+] \oplus x \\ &= \left[\bar{\nu}\mu', (\bar{\mu}'\mu' - \bar{\nu}\nu)\tfrac{1}{\sqrt{2}}, \bar{\mu}'\nu\right] \oplus \left((\bar{\mu}'\mu' + \bar{\nu}\nu)\tfrac{1}{\sqrt{2}}\right) \quad . \end{aligned} \tag{35}$$

The only other possibility is the more complex combination where the muon is a strongly masked $\mu$”. In addition to the coupling to $\nu$, since $\mu$” shows a strong $S^2$ electromagnetic interaction, $\mu$” also couples strongly to the electron $e$, forming two related SU(2) hadronic systems. We may substitute $\nu$ by $e$ as the leptonic doublet partner. We get the $K$ representation:

$$\begin{aligned} [\mu'', \nu] \otimes [\bar{\mu}'', \bar{\nu}] &\equiv [K^-, K_L, K^+] \oplus x_L' \\ &= \left[\bar{\nu}\mu'', (\bar{\mu}''\mu'' - \bar{\nu}\nu)\tfrac{1}{\sqrt{2}}, \bar{\mu}''\nu\right] \oplus \left((\bar{\mu}''\mu'' + \bar{\nu}\nu)\tfrac{1}{\sqrt{2}}\right) \quad , \end{aligned} \tag{36}$$

$$\begin{aligned} [\mu'', e] \otimes [\bar{\mu}'', \bar{e}] &\equiv [\bar{K}^0, K_s, K^0] \oplus x_S' \\ &= \left[\bar{e}\mu'', (\bar{\mu}''\mu'' - \bar{e}e)\tfrac{1}{\sqrt{2}}, \bar{\mu}''e\right] \oplus \left((\bar{\mu}''\mu'' + \bar{e}e)\tfrac{1}{\sqrt{2}}\right) \quad . \end{aligned} \tag{37}$$

The masses of these $\pi$ and $K$ geometric particles equal the masses of ground states belonging to the product representation of these representations. Their geometrical masses, essentially determined by the mass of its principal component or masked heavy lepton as indicated in equations (27) and (32), approximately correspond to the masses of all physical pions and kaons. We may also define

$$x' = \left(x'_L + x'_S\right)\tfrac{1}{\sqrt{2}} = \left(2\bar{\mu}''\mu'' + \bar{e}e + \bar{\nu}\nu\right)\tfrac{1}{2} \equiv \eta' \ . \tag{38}$$

$$\left(x + x'\right)\tfrac{1}{\sqrt{2}} \equiv \eta \ . \tag{39}$$

These definitions correspond to physical particles whose geometrical masses, essentially determined by the masked heavy lepton mass, are approximately,

$$m_{x'} \approx 990 \text{ Mev } + O(\alpha) \approx 957.8 \text{ Mev} = m_{\eta'} \tag{40}$$

and using the experimental value of $m_{\eta'}$,

$$m_x \approx 549 \text{ Mev } + O(\alpha) \approx 547.3 \text{ Mev} = m_\eta \ . \tag{41}$$

These results suggest a higher approximate symmetry for the $\pi K$ combination, that may be contained in the pseudoscalar meson representation. In other words, these mesons may be considered lepton-antilepton pairs, as suggested by Barut [11].

In addition to the proper topological $P$-excitations just discussed, we may consider ***subexcitations inside G***. For $[n=0]$ $P$-subexcitations, considered in $G$, there is a set of $C$ subspaces that correspond to the infinite ways of choosing $P\subset G$, not to the unique $C$ defined for an electron $P\subset P$. These additional $C$ spaces are not $P$-equivalent and must be included in the integration. For a triple electromagnetic interaction there are as many $C$ spaces as U(1) subgroups inside SU(2), in other words, as points on the sphere SU(2)/U(1). The volume of integration should be multiplied by $4\pi$. The same thing happens with $L$-excitations. In other words, class $[n=0]$ leptonic subexcitation masses $m_0$, when present in a hadronic $G$-system, are masked by the nonlinear strong interactions. We shall refer to these $S^2$ masked (strongly interacting) excitations as $G_S$-subexcitations. The $G_S$-subexcitations geometric mass values corresponding to the fundamental leptonic excitations are $4\pi$ times the previous leptonic geometric mass values,

$$m''\left(e, \nu_e, \mu, \nu_\mu, \tau, \nu_\tau\right) = \left(0.00641,\ 0,\ 1.77,\ 0,\ 29.9,\ 0\right) \quad \text{Gev} \ . \tag{42}$$

The masked excitations are characterized by the same quantum numbers that characterize the leptonic excitation but the value of the masked geometric mass includes the energy increase due to the other interactions. If the masked excitations were to be ejected (injected) from (to) a $G$-system they would loose (gain) the energy due to the extra interactions, and they would exit (enter) with the bare geometric mass value, as standard free leptons. There are no observable free particles with the masked masses. Experimentally the masked mass would never be detected by long-range methods. These masked leptonic subexcitations behave as quarks.

In particular, the $G_S$-subexcitations correspond, one to one, to the leptons. These $G_S$-subexcitations determine a hexadimensional space. Consequently any other possible geometric $G_S$-subexcitation can be expressed as a linear superposition of these fundamental masked geometric subexcitations. Alternately we may interpret some of these subexcitations as **quark** states. The structure of masked leptonic subexcitations is geometrically equivalent to the physical quark structure. For example, a superposition of 2 low velocity masked electrons $e$ plus 1 masked low energy neutrino $\nu_e$ has a $2/3$ charge and a total $4.2$ Mev invariant mass and may be interpreted as a $u$ quark. Similarly a superposition of 2 masked low velocity muons $\mu$ plus 1 masked low energy neutrino $\nu_\mu$ has $2/3$ charge and $1.2$ Gev total mass. It should be noted that, if free quarks are unobservable, all experimental information about their masses comes from bound quark states (meson resonances) and must depend somehow on theoretical arguments about these states. There is no quark confinement problem.

## 5. Conclusion.

The quotient for the geometric masses corresponding to $P$-excitations, due to the non linear substrate solution, including algebraic and topologic effects are approximately equal to the quotient of the bare masses of all known leptons. The necessary first order corrections, due to the interaction of the excitations, should be of order $\alpha$, equal to the order of the discrepancies.

The pion, kaon and other mesons may be considered as systems of two masked leptons. The mass of the pseudoscalar mesons may be explained as the mass of the ground states of the product representation of two masked fundamental lepton representations. The geometry determines the mass spectrum of geometric excitation ground states, which for low masses, essentially agrees with the physical particle mass spectrum. Quarks may be interpreted, in principle, as masked geometrical leptonic $P$ or $L$ subexcitations.

# 6. Appendix

### 6.1 Fiber Bundle

Essentially a fiber bundle is a projected space which localy is a product space. Define a fiber bundle in the following manner [12 ]. Consider

$$p{:}E \to M \quad (43)$$

an object of $S(Top)$, the arrow category of topological spaces (category formed by all morphisms of the category *Top* as objects and commutative diagrams as morphisms). If the mapping $p$ is surjective it is a projection, $E$ is the total space, $M$ is the base space and the triple is a projected space. If $m \in M$

$$p^{-1}(m) \equiv E_m \quad \text{vertical} . \quad (44)$$

The projected space is called a sheaf if the projection is a local homeomorphism. Each point $e$ of $E$ has an open neighborhood homeomorphic to an open neighborhood of $p(e)$.

Define a pseudocategory $B(M,F)$ taking as objects the products $U \times F$, where $U$ are the open subsets of $M$ and $F$ is a topological space, and as morphisms preserving the projection over $U$,

$$\alpha : U \times F \to V \times F$$

$$\mathrm{pr}_V \circ \alpha = \mathrm{id} \circ \mathrm{pr}_U \qquad U \cap V \neq \emptyset \ . \quad (45)$$

A fiber bundle is the manifold formed by a projected space $E$ over $M$ with an atlas valued in $B(M,F)$ compatible with the projection $p$. If the transition functions of the atlas form a group we obtain a fiber bundle with structure group $G$. If the fiber coincides with the structure group we have a principal fiber bundle $(E,M,G)$

### 6.2 Homotopic Product

There are topological excitations characterized by an integer number associated to the homotopy of group spaces, that we shall call winding, or more appropriately, wrapping numbers $n$ for the different homotopy groups. For a complete treatment of this subject see references [13 ,14].

The neighborhood of curves $C(Y,y_0)$ in a manifold $Y$ is the colection of all continuous mappings

$$f : I^1 \to Y \quad (46)$$

of the unit interval that satisfy

$$f(0) = y_0 = f(1) \ . \quad (47)$$

Let $f$ and $g$ be two mappings in $C(Y,y_0)$. The juxtaposition of $f$ and $g$ is the element of $C(Y,y_0)$ given by

$$\begin{aligned}(f * g)(x) &= f(2x) \qquad \left(0 \le x \le 1/2\right) \\ &= g(2x) \qquad \left(1/2 \le x \le 1\right) \ . \end{aligned} \quad (48)$$

Similarly consider the mappings from the $n$-cube into $Y$

$$f : I^n \to Y \quad (49)$$

such that they send the boundary of $I^n$ to point $y_0$ of $Y$. The boundary $\beta I^n$ of $I^n$ is defined as the points that satisfy

$$\prod_{i=1}^{n} x_i (1 - x_i) = 0 \ . \quad (50)$$

That is, $f$ satisfies

$$f(\beta I^n) = y_0 \ . \quad (51)$$

Let us denote the set of these mappings by $C_n(Y,y_0)$.

Define a homotopy relation in $C_n(Y,y_0)$ saying that $f$ and $g$ are homotopic modulus $y_0$ if there is a continuous mapping

$$h : I^n \times I^1 \to Y \tag{52}$$

that satisfies

$$h(x,0) = f(x) \qquad x \in I^n \quad , \tag{53}$$

$$h(x,1) = g(x) \qquad x \in I^n \quad , \tag{54}$$

$$h(\beta I^n, t) = y_0 \qquad 0 \le t \le 1 \quad . \tag{55}$$

This is an equivalence relation in $C_n(Y,y_0)$ and decomposes it into equivalence classes formed by the arcwise connected components of $C_n(Y,y_0)$

La juxtaposition of two elements of $C_n(Y,y_0)$ is the element given by

$$\begin{aligned}(f*g)(x) &= f(2x_1, x_2, \cdots x_n) \qquad \left(0 \le x_1 \le \tfrac{1}{2}\right) \\ &= g(2x_1 - 1, x_2, \cdots x_n) \qquad \left(\tfrac{1}{2} \le x_1 \le 1\right) \quad .\end{aligned} \tag{56}$$

The homotopy group $\pi_n$ of $Y$ on the point $y_0$ is defined as those elements which are arcwise connected components of $C_n(Y,y_0)$ with group operation

$$[f] \circ [g] = [f*g] \quad . \tag{57}$$

These definitions, based on the mappings from the $n$-cube into $Y$, may be expressed as mappings from spheres

$$h : S^n \to M \quad . \tag{58}$$

### *6.3 Third Homotopy Group*

In particular we are interested in the third homotopy group of group spaces. To determine $\pi_3(G)$, we recognize that the exponential mapping from the maximal non compact subalgebra of sl(4,R) is a diffeomorfism [15] onto the coset, a non compact Riemannian subspace which is contractile. We have an exact short sequence in the general homotopy sequence [14],

$$\cdots \pi_4\left(G/H\right) \xrightarrow{\Delta_*} \pi_3(H) \xrightarrow{i_*} \pi_3(G) \xrightarrow{p_*} \pi_3\left(G/H\right) \cdots \quad , \tag{59}$$

$$\{0\} \longrightarrow \pi_3(H) \xrightarrow{i_*} \pi_3(G) \longrightarrow \{0\} \quad , \tag{60}$$

that implies the intermediate mapping is an isomorphism and

$$\pi_3(G) = \pi_3(H) \tag{61}$$

where $H$ is the maximal compact subgroup. It is known also that there is an isomorphism between the homotopy groups of a group and its covering group, except for the first homotopy group [16 ]. For the homotopy group of SL(4,R) we obtain

$$\begin{aligned}\pi_3(SL(4,R)) &= \pi_3(SU(2) \otimes SU(2)) \\ &= \pi_3(SU(2)) \otimes \pi_3(SU(2)) = Z \otimes Z \quad ,\end{aligned} \tag{62}$$

Similarly, we have for the homotopy groups of the other two possible holonomy groups,

$$\pi_3(Sp(4,R)) = \pi_3(R \otimes SU(2)) = \pi_3(SU(2)) = Z \quad , \tag{63}$$

$$\pi_3(SL(2,C)) = \pi_3(SU(2)) = Z \quad . \tag{64}$$

### 6.4 Volumes.

There is a constant solution [17 ] for the differential equations that determine the substratum connection. This local solution may generate global scattering solutions with different topology in accordance with the third homotopy group of the transition functions of the manifold. The excitations, although locally defined around a local trivial section (solution), are related by transition functions to associated global topological solutions, characterized by an integer number called wrapping number $n$.

The integrand $J\bullet\Gamma$ is a local constant, equal for all $n$ values. The integration is on a subspace $K_R \subset K$ of relativistically inequivalent points of $K$ for the group $G$, and on a subspace $C_R \subset C \subset K$ for a subgroup $H$. When integrating over local trivial states $n=0$, the expressions for the masses are [17],

$$m_G = \frac{m_g}{4V\left(A_R\right)} \operatorname{tr} \int_{K_R} F(k,k)dk = \frac{V\left(K_R\right)}{4V\left(A_R\right)} m_g \operatorname{tr}\left(F_0\right) \quad , \tag{65}$$

$$m_H = \frac{m_g}{4V\left(A_R\right)} \operatorname{tr} \int_{C_R} F(k,k)dk = \frac{m_g}{4V\left(A_R\right)} \operatorname{tr}\left(F_0\right) V\left(C_R\right) = \frac{m_G V\left(C_R\right)}{V\left(K_R\right)} \quad . \tag{66}$$

The volumes of the three cosets ($K$, $C$, $I$ ) determined by the chain $G \supset H \supset L$ were previously calculated [17, 18]. It is necessary to eliminate the equivalent states dividing by the equivalence relation $R$ under SO(3,1) boosts. Equivalent points are related by a Lorentz boost transformation of magnitude $\beta$. There are as many equivalent points as the volume of the orbit $R$ developed by parameter $\beta$. The respective inequivalent volumes are,

$$V\left(K_R\right) = \frac{V(K)}{V\left(R(\beta)\right)} = \frac{2^5 \pi^6 I_K(\beta)}{I_K(\beta)} = 2^5 \pi^6 \quad . \tag{67}$$

$$V\left(C_R\right) = \frac{V(C)}{V\left(R(\beta)\right)} = \frac{\frac{16\pi}{3} I_C(\beta)}{I_C(\beta)} = \frac{16\pi}{3} \quad , \tag{68}$$

$$V_R\left(L/L\right) = V(I) = 0 \quad . \tag{69}$$

References

1 G. González-Martín, Gen. Rel. Grav. 26, 1177 (1994).
2 G. González-Martín, Phys. Rev. D35, 1225 (1987).
3 E. Mach, The Science of Mechanics, 5th English ed. (Open Court, LaSalle), ch. 1 (1947).
4 A. Einstein The Meaning of Relativity, 5th ed. (Princeton Univ. Press, Princeton), p.55 (1956).
5 F. Lenz, Phys. Rev. 82, 554 (1951).
6 I. J. Good, Phys. Lett. 33A, 383 (1970).
7 A. Wyler, Acad, Sci. Paris, Comtes Rendus, 271A, 180 (1971).
8 G. González-Martín, Rev. Mex. Fis. 49, Sup. 3, 120 (2003).
9 R. E. Marshak, Conceptual Foundations of Modern Particle Physics, (World Scientific, Singapore) ch. 10 (1993).
10 R. Hermann, Lie Groups for Physicists (W. A. Benjamin, New York) p. 53 (1966).
11 A. O. Barut, Surv. High Energy Phys.1, 113 (1980).
12 I. Vaisman, Cohomology of Differential Forms, (Marcel Dekker, New York), p. (1973).
13 J. G. Hocking, G.S. Young, Topology (Addison-Wesley, Reading) p. 159 (1961).
14 G. W. Whitehead, Elements of Homotoy Theory (Springer Verlag, New York) (1978).
15 S. Helgason, Differential Geometry and Symmetric Spaces (Academic Press, New York) p. 130, 137, 214 (1962).
16 F. H. Croom, Basic Concepst in Algebraic Topology, (Springer Verlag, New York) (1978).
17 G. González-Martín, Report SB/F/274-99; lanl archive physics/0009066 (2000).
18 G. González-Martín, Physical Geometry, (Universidad Simón Bolívar, Caracas) (2000).